# Collection of Master-Slave Synchronized Chaotic Systems

**A.I.Lerescu[a], N.Constandache[b], S.Oancea[c], I.Grosu[b]**

[a] *Physics Department ,Groningen University, The Netherlands*

[b] *Faculty of Bioengineering, University of Medicine and Pharmacy, Iasi, 6600, Romania*

[c] *Biophysics Dept. ,University of Agriculture and Veterinary Medicine, Iasi 6600, Romania*

## Abstract

In this work the Open-Plus-Closed-Loop (OPCL) method of synchronization is used in order to synchronize the systems from the Sprott's collection of the simplest chaotic systems. The method is general and we looked for the simplest coupling between master and slave .The main result is that for the systems that contains one nonlinear term and that term contains one variable then the coupling consists of one term. The numerical intervals of parameters where the synchronization is achieved are obtained analytically by applying Routh-Hurwitz conditions. Detailed calculations and numerical results are given for the system I from the Sprott’s collection. Working in the same manner for many systems this method can be adopted for the teaching of the topic.

## 1.Introduction

Synchronization of chaotic systems has received a significant attention since 1990.The state of the art is contained in two books [1,2] and a review paper [3] and several special issues of some journals. There are known different types of synchronization: mutual, master-slave, weak, strong, phase and generalized.

Biological cells are complex multivariable systems. How they are connected to do a specific task is a challenge for the today science [1]. A huge effort is being spent to obtain experimental data. These data need good models to get a deep understanding. How the models are used to achieve synchronization knows at least two opposite tendencies: one is to propose an intuitive coupling and to find numerically the feedback gains. Another one is to propose a coupling that assures the synchronization and to look for a justification of the coupling. The coupling can be realized by modifying a parameter [4] or by using an additive coupling [5].

Another additive coupling is offered by the open-plus-closed-loop (OPCL) method of control [6]. OPCL synchronization [7] uses the Taylor expansion for obtaining a general coupling term. This strategy has been used and improved and modified [8]. Very recently other methods are proposed [9-14].

Ten years ago, Sprott [15] proposed a collection of 19 chaotic systems that are simpler than Lorenz’s and Rossler’s systems. Sprott’s systems have either five terms and two nonlinearities or six terms and one nonlinearity. Lorenz’s system has seven terms and two nonlinearities and Rossler’s system has seven terms and one nonlinearity.

In this paper we show that OPCL synchronization [7] can be applied to all 19 systems from the Sprott's collection[15]. For every master system we give the slave system .The method can be applied to any other systems. The advantages and disadvantages are discussed. We believe that this method could be adopted for the teaching of the topic.

The paper is organized as follows. Section 2 contains the details of the OPCL synchronization. All calculations and numerical results are given for one system in Section 3. Also here the results for the other systems are presented in Table I. A Liapunov function has been found for one of the systems. The final remarks and conclusions are presented in Section 4.

## 2.OPCL Method of Synchronization

The original OPCL [6] offers a driving for a general system:

$$d\mathbf{x}/dt=\mathbf{F}(\mathbf{x}) \ ; \quad \mathbf{x}\in \mathbf{R}^n \qquad (1)$$

in order to achieve a desired goal dynamics $\mathbf{g}(t) \in \mathbf{R}^n$ . The driving has two terms :$\mathbf{D}_1(\mathbf{g})$ as an open-loop driving and $\mathbf{D}_2(\mathbf{x},\mathbf{g})$ as an closed-loop (feedback ) one.

$$\mathbf{D}(\mathbf{x},\mathbf{g}) = \mathbf{D}_1(\mathbf{g}) + \mathbf{D}_2(\mathbf{x},\mathbf{g}) \qquad (2)$$

with:

$$\mathbf{D}_1(\mathbf{g}) = d\mathbf{g}/dt-\mathbf{F}(\mathbf{g}) \qquad (3)$$

$$\mathbf{D}_2(\mathbf{x},\mathbf{g})= (\mathbf{H}-\partial\mathbf{F}(\mathbf{g})/\partial\mathbf{g}\,)\ (\mathbf{x}-\mathbf{g}) \qquad (4)$$

where $\mathbf{H}$ is an arbitrary constant Hurwitz matrix . The driven system :

$$d\mathbf{x}/dt=\mathbf{F}(\mathbf{x}) + \mathbf{D}(\mathbf{x},\mathbf{g}) \qquad (5)$$

has an asymptotic behavior $\mathbf{x}(t) \rightarrow \mathbf{g}(t)$ , if $\| \mathbf{x}(0)-\mathbf{g}(0) \|$ is small enough [6,7]. This result has a simple explanation: with $\mathbf{e}=\mathbf{x}-\mathbf{g}$ and taking the Taylor expansion $\mathbf{F}(\mathbf{x})=\mathbf{F}(\mathbf{g}+\mathbf{e})=\mathbf{F}(\mathbf{g})+ (\partial\mathbf{F}(\mathbf{g})/\partial\mathbf{g}\,)\mathbf{e} +\ldots$ from (5),(4),(2),(3) follows:

$d\mathbf{e}/dt=\mathbf{He}$ that assures that $\mathbf{e} \rightarrow \mathbf{0}$.

This strategy has been applied to master-slave synchronization [7], using the master's dynamics as the goal dynamics for the slave system :$\mathbf{g}(t)=\mathbf{X}(t)$. For a master system

$$d\mathbf{X}/dt=\mathbf{F}(\mathbf{X}) \ ; \qquad \mathbf{X}\in \mathbf{R}^n \qquad (6)$$

the slave system is :

$$d\mathbf{x}/dt=\mathbf{F}(\mathbf{x}) + \mathbf{D}(\mathbf{x},\mathbf{X}) \qquad (7)$$

with:

$$\mathbf{D}(\mathbf{x},\mathbf{X}) = \mathbf{D}_2(\mathbf{x},\mathbf{X}) = (\mathbf{H}- \partial\mathbf{F}(\mathbf{X})/\partial\mathbf{X}\,)(\mathbf{x}-\mathbf{X}) \qquad (8)$$

because $\mathbf{D}_1(\mathbf{X})$ is zero.

The driving term (8) is general and can be more or less complicated depending on the structure of the system. The matrix **H** is an arbitrary constant Hurwitz matrix that can be chosen in such a manner that (8) to be as simple as possible.

The simplest $\mathbf{D}_2(\mathbf{x},\mathbf{X})$ term is one that has all elements zero except one $(\mathbf{D}_2(\mathbf{x},\mathbf{X}))_{ik}$ = constant $(x_k\text{-}X_k)$ that should be added in equation (i) of the system (7).

This is not possible because $\partial\mathbf{F}(\mathbf{X})/\partial\mathbf{X}$ has variable terms. When $(\partial\mathbf{F}(\mathbf{X})/\partial\mathbf{X})_{ik}$ is constant we can choose $H_{ik}= (\partial\mathbf{F}(\mathbf{X})/\partial\mathbf{X})_{ik}$ and $(\mathbf{D}_2(\mathbf{x},\mathbf{X}))_{ik}$ will be zero. We need to introduce a parameter p (or several parameters ) in the matrix **H** and to look for the values of this(these) parameter(s) that assure that **H** is Hurwitz matrix( see next section).

From this we can conclude that $\mathbf{D}_2(\mathbf{x},\mathbf{X})$ will be simpler if $\mathbf{F}(\mathbf{X})$ has fewer nonlinear terms. The simplest case will be when $\mathbf{F}(\mathbf{X})$ has just one nonlinear term that contains just one variable. In this case $\partial\mathbf{F}(\mathbf{X})/\partial\mathbf{X}$ has one variable term. Let's consider it as the term (i,k). We choose $H_{ik}=p$ , where p is a parameter that should be determined in order that **H** to be a Hurwitz matrix. If the Routh-Hurwitz can not be satisfied with one parameter then we are forced to introduce a second one(see below, Table I, the systems B,D,E).

## 3.Applications

We give detailed calculations for the system I from Sprott's collection (see also Table I ,below). The master system has the form:

$$dX_1/dt =-0.2X_2\,;\ dX_2/dt=X_1 + X_3\,;dX_3/dt =X_1 -X_3 +X_2^2 \qquad (9)$$

In this case

$$\partial\mathbf{F}(\mathbf{X})/\partial\mathbf{X} = \begin{pmatrix} 0 & -0.2 & 0 \\ 1 & 0 & 1 \\ 1 & 2X_2 & -1 \end{pmatrix} \qquad (10)$$

We can choose

$$\mathbf{H} = \begin{pmatrix} 0 & -0.2 & 0 \\ 1 & 0 & 1 \\ 1 & p & -1 \end{pmatrix} \qquad (11)$$

The characteristic equation of (11) is:

$$\lambda^3 +a_1\lambda^2+ a_2\lambda+a_3=0 \qquad (12)$$

with

$$a_1=1\ ;\ a_2= -p+0.2\ ;\ a_3=0.4 \qquad (13)$$

The Routh-Hurwitz conditions (for characteristic equation (12) to have negative real part eigenvalues ) are :

$$a_1 > 0 \ ; \ a_1a_2 - a_3 > 0 \ ; \qquad a_3 > 0 \qquad (14)$$

The conditions (14) for (13) give $-0.2>p$.

So the slave system (of the master system (9)) is:

$$dx_1/dt=-0.2x_2 \ ; dx_2/dt=x_1+x_3; \ dx_3/dt=x_1-x_3+x_2^2 \ +(p-2X_2)(x_2-X_2) \qquad (15)$$

with $-0.2>p$.

This result is in Table I (below) along with the results for the other systems.

The above coupling is the simplest possible. It is necessary to transmit the variable $X_2$ from the master and to add the coupling term $(p-2X_2)(x_2-X_2)$ in the 3-rd equation of the slave system.

Numerical results are given in Fig. 1 for the system (I) master system (9).

The results for all 19 systems from Sprott's collection are presented in Table I that contains a column with the master systems, a column with the slave system and a column with the values of the parameter (parameters). For the systems F,H,I,J,L,M,N,P.Q,S the coupling is the simplest : one parameter and one term. For the systems G,K,O,R the coupling contains one parameter and two terms. For the system C we need one parameter and three terms. The worst situation is for the system E: two parameters and five terms.

We should mention that OPCL meant as a linear approximation because we take just only linear term in Taylor expansion. It means that the distance $\| \mathbf{x}(0)-\mathbf{X}(0) \|$ should be small enough. In [7] we found a Liapunov function for the Lorenz system. If a Liapunov function can be found then the synchronization can be obtained for any **x**(0) and **X**(0).

Let's consider the system (A):

$$dX_1/dt=X_2 \ ;$$
$$dX_2/dt= -X_1+X_2X_3 \ ; \qquad (16)$$
$$dX_3/dt=1-X_2^2$$

Using the above presented algorithm we obtained a rather complicated coupling and a narrow interval for the parameter p(see Table I). Using another coupling with **H** and $\partial\mathbf{F(X)}/\partial\mathbf{X}$:

$$\mathbf{H}=\begin{pmatrix} 0 & 1 & 0 \\ -1 & p & 0 \\ 0 & 0 & 0 \end{pmatrix} \qquad \partial\mathbf{F(X)}/\partial\mathbf{X}=\begin{pmatrix} 0 & 1 & 0 \\ -1 & X_3 & X_2 \\ 0 & -2X_2 & 0 \end{pmatrix} \qquad (17)$$

and **e**= **x**-**X** we have :

$$de_1/dt= e_2 \ ;de_2/dt= -e_1+pe_2+e_2e_3 \ ;de_3/dt=-e_2^2 \qquad (18)$$

Choosing the Liapunov function

$$L= (e_1^2+e_1^2+e_3^2)/2 \qquad (19)$$

we have $dL/dt=pe_2^2$ and $dL/dt<0$ for $p<0$.

So for the system A we can have also the slave system:

$$dx_1/dt=x_2;$$
$$dx_2/dt=-x_1+x_2x_3+(p-X_3)(x_2-X_2)-X_2(x_3-X_3)\ ; \qquad (20)$$
$$dx_3/dt=1-x_2^2+2X_2(x_2-X_2)$$

This coupling is simpler because we need to record and transmit 2 variables $X_2$ and $X_3$. In addition in this case the basin of entrainment is the whole space, it means that the synchronization can be achieved starting from any initial conditions **x**(0) and **X**(0). Unfortunately we were not able to find Liapunov functions for the other systems.

## 4. Conclusions

In this paper we applied the OPCL synchronization[7] to the simplest chaotic systems that are known as the Sprott's collection[15]. This collection contains 19 chaotic systems and we give the coupling master-slave for all systems from the collection. We tried to find as simple as possible coupling. The simplest one contain just one term and a variable coefficient. The algorithm is general and straightforward and we think that it could be adopted for the teaching of the topic.

## References

[1]Mosekilde E,Mastrenko Y,Postnov D .Chaotic synchronization.Applications for Living Systems.Singapore:World Scientific,2002

[2] Pikovsky A,Rosenblum M,Kurths J.Synchronization,a Universal Concept in Nonlinear Science.Cambridge: Cambridge Univ. Press,2001

[3] Boccaletti S, The synchronization of chaotic systems Physics Reports 2002;366:1-101

[4] Liu Zonghua, Chen Shigang.General method of synchronization.Phys. Rev. E 1997;55:6651-6655

[5]Ali MK, Fang JQ Synchronization of chaos and hyperchaos using linear and nonlinear feedback functions. Phys. Rev. E 1997;55:5285-5290

[6]Jackson EA,Grosu I An open-plus-closed-loop(OPCL) control of complex dynamic systems.Physica D 1995;85:1-9
[7]Grosu I Robust synchronization. Phys. Rev. E 1997;56:3709-3712
[8]Chen L-Q ,Liu Y-Z An open-plus-closed-loop approach to synchronization of chaotic and hyperchaotic maps .Int. J. of Bif. and Chaos 2002;12(5):1219-1225
[9]Jiang GP,Tang KS,Chen GR.A simple global synchronization criterion for coupled chaotic systems .Chaos,Solitons &Fractals 2003;15:925-935
[10]Chen S,Wang F,Wang C.Synchronizing strict-feedback and general strict-feedback chaotic systems via a single controller.Chaos,Solitons &Fractals 2004;20:235-243
[11] Sun JT, Zang YP. Some simple global synchronization criterions for coupled time-varying chaotic systems Chaos,Solitons &Fractals 2004 ;19:789-794
[12] Wang CC, Su JP .A new adaptive variable structure control for chaotic synchronization and secure communication. Chaos,Solitons &Fractals 2004;20:967-977
[13]Chen S,Zhang Q, Xie J, Wang C. A stable-manyfold method for chaos control and synchronization . Chaos,Solitons &Fractals 2004;20:947-954
[14]Huang L,feng R,wang M. Synchronization of chaotic systems via nonlinear control.Phys. Lett. A 2004;320:271-275
[15]Sprott JC.Some simple chaotic flows. Phys. Rev.E 1994;55(5):5285-5290

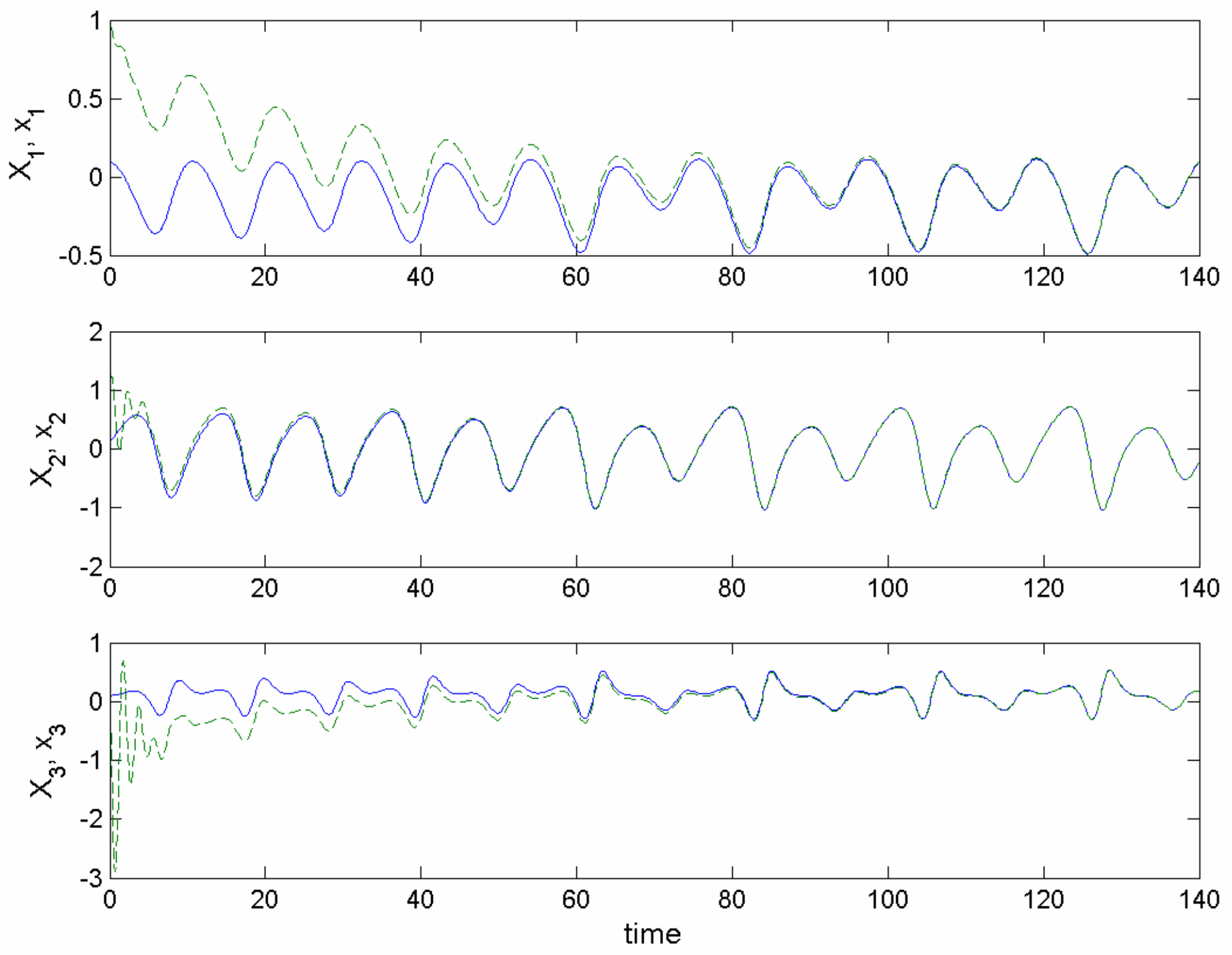

a)

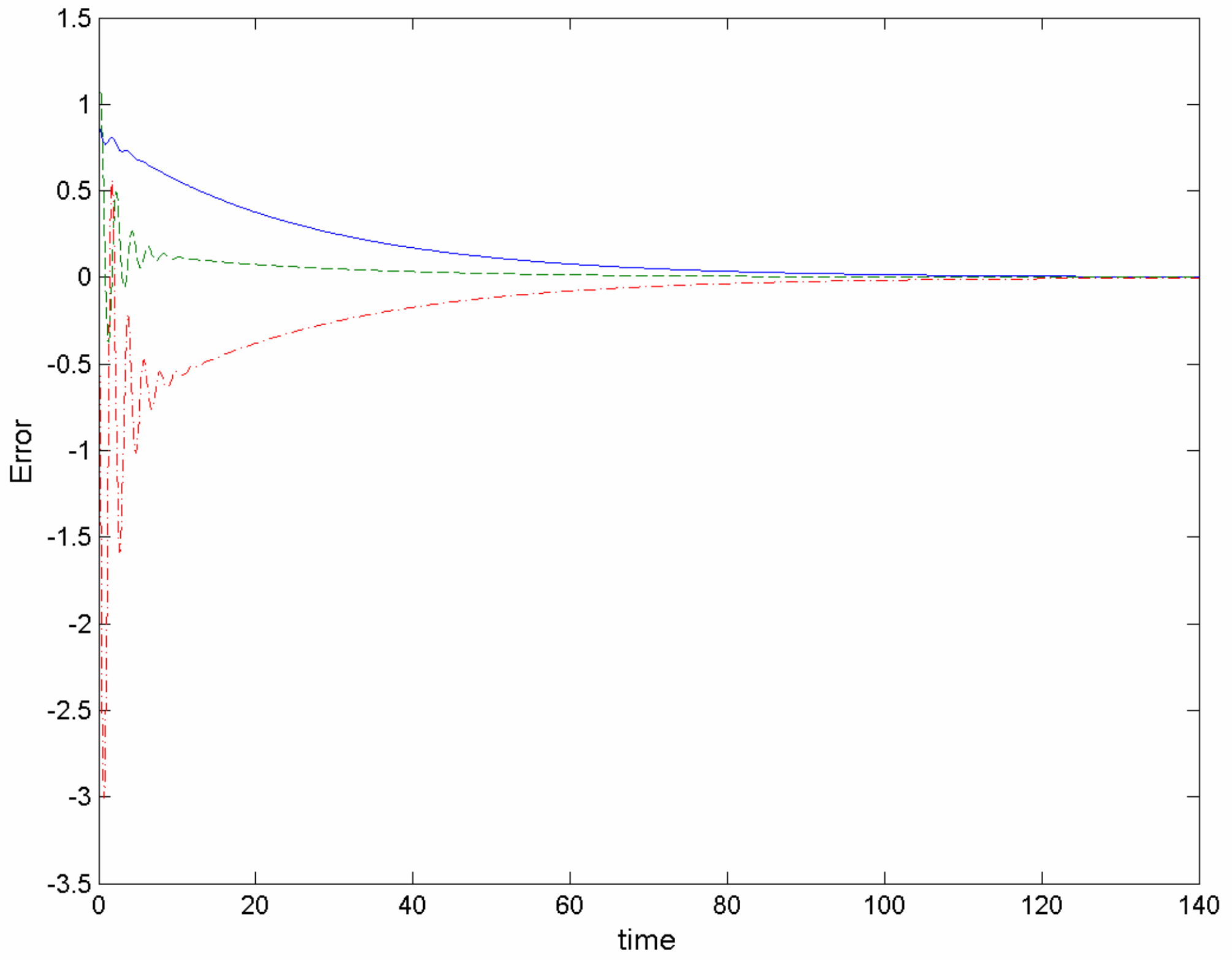

b)

**Figure 1**

**Figure Captions**

**Figure 1.** a) Numerical results for system (I) for $p = -10$ and $X_1(0) = X_2(0) = X_3(0) = 0.1$ and $x_1(0) = x_2(0) = x_3(0) = 1$ and b) time dependence of the error.

**Table I.** Sprott's collection of the simplest chaotic flows (master systems) and the corresponding slave systems (slave) . Last column shows the values of the control parameters.

## Table I

| System | Master | Slave | Parameter |
|---|---|---|---|
| A | $dX_1/dt= X_2$<br>$d X_2/dt=- X_1 + X_2 X_3$<br>$d X_3/dt=1- X_2^2$ | $d x_1/dt= x_2$<br>$d x_2/dt=- x_1+ x_2 x_3+(-1- X_3)( x_2- X_2)+(1- X_2)(x_3- X_3)$<br>$d x_3/dt=1- x_2^2+p(x_1-X_1)+2 X_2(x_2- X_2)$ | p=(-1,0) |
| B | $d X_1/dt= X_2 X_3$<br>$d X_2/dt= X_1 - X_2$<br>$d X_3/dt=1- X_1 X_2$ | $d x_1/dt= x_2 x_3+(1- X_3)( x_2- X_2)+(p1- X_2)( x_3- X_3)$<br>$d x_2/dt= x_1- x_2$<br>$d x_3/dt=1- x_1 x_2+(p2+ X_2)( x_1- X_1 )+(1+X_1)( x_2- X_2)$ | $p1>1$<br>$p2<-1$ |
| C | $d X_1/dt= X_2 X_3$<br>$d X_2/dt= X_1 - X_2$<br>$d X_3/dt=1- X_1^2$ | $d x_1/dt= x_2 x_3+(p- X_3)( x_2- X_2)-(1+ X_2)( x_3- X_3)$<br>$d x_2/dt= x_1- x_2$<br>$d x_3/dt=1- x_1^2+(1+2 X_1)( x_1- X_1)$ | $p<0$ |
| D | $d X_1/dt=- X_2$<br>$d X_2/dt= X_1 + X_3$<br>$d X_3/dt= X_1 X_3+3 X_2^2$ | $d x_1/dt=- x_2$<br>$d x_2/dt= x_1+ x_3$<br>$d x_3/dt= x_1 x_3+3 x_2^2+(p2- X_3)( x_1- X_1 )-6 X_2(x_2- X_2)+(p1- X_1 )( x_3- X_3)$ | $p1<0$<br>$p1<p2$<br>$p2<0$ |
| E | $d X_1/dt= X_2 X_3$<br>$d X_2/dt= X_1^2- X_2$<br>$d X_3/dt=1-4 X_1 X_2$ | $d x_1/dt= x_2 x_3+(p1- X_3)( x_2- X_2)+(p2- X_2)( x_3- X_3)$<br>$d x_2/dt=x_1^2- x_2+(1-2 X_1)( x_1- X_1)$<br>$d x_3/dt=1-4 x_1 x_2+(-4+4X_2)(x_1-X_1)+4X_1(x_2-X_2)$ | $p1<0$<br>$p2>0$ |
| F | $d X_1/dt= X_2+ X_3$<br>$d X_2/dt=- X_1 +0.5 X_2$<br>$d X_3/dt= X_1^2- X_3$ | $d x_1/dt= x_2+ x_3$<br>$d x_2/dt=- x_1+0.5 x_2$<br>$d x_3/dt= x_1^2-x_3+(p-2 X_1)( x_1- X_1)$ | p=(-2;-0.75) |
| G | $d X_1/dt=0.4 X_1 + X_3$<br>$d X_2/dt= X_1 X_3- X_2$<br>$d X_3/dt=- X_1 + X_2$ | $d x1/dt=0.4 x_1+ x3$<br>$d x2/dt= x_1 x_3- x_2- X_3 (x- X_1)+(p- X_1)( x_3- X_3)$<br>$d x3/dt=- x_1+ x_2$ | p=(-2.5;-0.6) |
| H | $d X_1/dt=- X_2+ X_3^2$<br>$d X_2/dt= X_1 +0.5 X_2$<br>$d X_3/dt= X_1 - X_3$ | $d x_1/dt=- x_2+ x_3^2+(p-2 X_3)( x_3- X_3)$<br>$d x_2/dt= x_1+0.5 x_2$<br>$d x_3/dt= x_1- x_3$ | p=(-2;-0.75) |
| I | $d X_1/dt=-0.2 X_2$<br>$d X_2/dt= X_1 + X_3$<br>$d X_3/dt= X_1 + X_2^2- X_3$ | $d x_1/dt=-0.2 x_2$<br>$d x_2/dt= x_1+ x_3$<br>$d x_3/dt= x_1+ x_2^2-x_3+(p-2 X_2)( x_2- X_2)$ | $p<-0.2$ |
| J | $d X_1/dt=2 X_3$<br>$d X_2/dt=-2 X_2+ X_3$<br>$d X_3/dt=- X_1 + X_2+ X_2^2$ | $d x_1/dt=2 x_3$<br>$d x_2/dt=-2 x_2+ x_3$<br>$d x_3/dt=- x_1+x_2+x_2^2+(p-1-2X_2)( x_2- X_2)$ | $p<0$ |
| K | $d X_1/dt= X_1 X_2- X_3$<br>$d X_2/dt= X_1 - X_2$<br>$d X_3/dt= X_1 +0.3 X_3$ | $d x_1/dt= x_1x_2- x_3-X_2(x_1- X_1)+(p- X_1)( x_2- X_2)$<br>$d x_2/dt= x_1- x_2$<br>$d x_3/dt= x_1+0.3 x_3$ | p=(-3.3;-0.51) |
| L | $d X_1/dt= X_2+3.9 X_3$<br>$d X_2/dt=0.9 X_1^2- X_2$<br>$d X_3/dt=1- X_1$ | $d x_1/dt= x_2+3.9 x_3$<br>$d x_2/dt=0.9 x_1^2- x_2+(p-1.8 X_1)( x_1- X_1)$<br>$d x_3/dt=1- x_1$ | $p<0$ |
| M | $d X_1/dt=- X_3$<br>$d X_2/dt=- X_1^2- X_2$<br>$d X_3/dt=1.7+1.7 X_1 +X_2$ | $d x_1/dt=- x_3$<br>$d x_2/dt=- x_1^2- x_2+(p+2 X_1)( x_1- X_1)$<br>$d x_3/dt=1.7+1.7 x_1+ x_2$ | p=(-1.7,0) |
| N | $d X_1/dt=-2 X_2$<br>$d X_2/dt= X_1 + X_3^2$<br>$d X_3/dt=1+ X_2-2 X_3$ | $d x_1/dt=-2 x_2$<br>$d x_2/dt= x_1+ x_3^2+(p-2 X_3)( x_3- X_3)$<br>$d x_3/dt=1+ x_2-2 x_3$ | $p<0$ |
| O | $d X_1/dt= X_2$<br>$d X_2/dt= X_1 - X_3$<br>$d X_3/dt= X_1 + X_1 X_3+2.7X_2$ | $d x_1/dt= x_2$<br>$d x_2/dt= x_1- x_3$<br>$d x_3/dt= x_1+ x_1x_3+2.7 x_2- X_3(x_1- X_1)+(p- X_1)( x_3-X_3)$ | p=(-1 ;-0.37) |
| P | $d X_1/dt=2.7 X_2+ X_3$<br>$d X_2/dt=- X_1 + X_2^2$<br>$d X_3/dt= X_1 + X_2$ | $d x_1/dt=2.7 x_2+ x_3$<br>$d x_2/dt=- x_1+ x_2^2+(p-2 X_2)( x_2- X_2)$<br>$d x_3/dt= x_1+ x_2$ | p=(-1;-10/27) |
| Q | $d X_1/dt=- X_3$<br>$d X_2/dt= X_1 - X_2$<br>$d X_3/dt=3.1 X_1 +X_2^2+0.5 X_3$ | $d x_1/dt=- x_3$<br>$d x_2/dt= x_1- x_2$<br>$d x_3/dt=3.1 x_1+ x_2^2+0.5 x_3+(p-2 X_2)( x_2- X_2)$ | p=(-3.1;-1.8) |
| R | $d X_1/dt=0.9- X_2$<br>$d X_2/dt=0.4+ X_3$<br>$d X_3/dt= X_1 X_2- X_3$ | $d x_1/dt=0.9- x_2$<br>$d x_2/dt=0.4+ x_3$<br>$d x_3/dt= x_1 x_2- x_3+(1- X_2)( x_1- X_1)+(p- X_1)( x_2- X_2)$ | $p<-1$ |
| S | $d X_1/dt=- X_1 -4 X_2$<br>$d X_2/dt= X_1 + X_3^2$<br>$d X_3/dt=1+ X_1$ | $d x_1/dt=- x_1-4 x_2$<br>$d x_2/dt= x_1+x_3^2+(p-2 X_3)(x_3- X_3)$<br>$dz/dt=1+ x_1$ | p=(0,1) |